\documentclass[aps,prl,reprint,showpacs,bibnotes,superscriptaddress]{revtex4-1}

\usepackage[dvipdfmx]{graphicx}
\usepackage{amssymb}
\usepackage{color}
\begin{document}

\title{Spontaneous Hall effect induced by strain in Pr$_2$Ir$_2$O$_7$ epitaxial thin films}

\author{Takumi~Ohtsuki}
\email[Author to whom correspondence should be addressed.\\ E-mail: ]{ohtsuki@issp.u-tokyo.ac.jp}
\affiliation{Institute for Solid State Physics, The University of Tokyo, Kashiwa, Chiba 277-8581, Japan}
\author{Zhaoming~Tian}
\affiliation{Institute for Solid State Physics, The University of Tokyo, Kashiwa, Chiba 277-8581, Japan}
\author{Akira~Endo}
\affiliation{Institute for Solid State Physics, The University of Tokyo, Kashiwa, Chiba 277-8581, Japan}
\author{Mario~Halim}
\affiliation{Institute for Solid State Physics, The University of Tokyo, Kashiwa, Chiba 277-8581, Japan}
\author{Shingo~Katsumoto}
\affiliation{Institute for Solid State Physics, The University of Tokyo, Kashiwa, Chiba 277-8581, Japan}
\author{Yoshimitsu~Kohama}
\affiliation{Institute for Solid State Physics, The University of Tokyo, Kashiwa, Chiba 277-8581, Japan}
\author{Koichi~Kindo}
\affiliation{Institute for Solid State Physics, The University of Tokyo, Kashiwa, Chiba 277-8581, Japan}
\author{Satoru~Nakatsuji}
\affiliation{Institute for Solid State Physics, The University of Tokyo, Kashiwa, Chiba 277-8581, Japan}
\affiliation{CREST, Japan Science and Technology Agency (JST), 4-1-8 Honcho Kawaguchi, Saitama 332-0012, Japan}
\author{Mikk~Lippmaa}
\affiliation{Institute for Solid State Physics, The University of Tokyo, Kashiwa, Chiba 277-8581, Japan}

\date{\today}

\begin{abstract}
Strongly correlated iridate pyrochlores with geometrically frustrated spins have been recognized as a potentially interesting group of oxide materials where novel topological phases may appear. A particularly attractive system is the metallic Pr$_2$Ir$_2$O$_7$, as it is known as a Fermi node semimetal characterized by quadratic band touching at the Brillouin zone center, suggesting that the topology of its electronic states can be tuned by moderate lattice strain. In this work we report the growth of epitaxial Pr$_2$Ir$_2$O$_7$ thin films grown by solid-state epitaxy. We show that the strained parts of the films give rise to a spontaneous Hall effect that persists up to 50 K without having spontaneous magnetization within our experimental accuracy. This indicates that a macroscopic time reversal symmetry (TRS) breaking appears at a temperature scale that is too high for the magnetism to be due to Pr 4$f$ moments, and must thus be related to magnetic order of the iridium 5$d$ electrons. The magnetotransport and Hall analysis results are consistent with the formation of a Weyl semimetal state that is induced by a combination of TRS breaking and cubic symmetry breaking due to lattice strain.
\end{abstract}

%\pacs{73.50.-h, 75.47.-m, 75.30.Mb}
%\keywords{}

\maketitle

\section{Introduction}
The search for novel topological phases in strongly correlated electron systems is one of the frontiers of modern condensed matter physics. In this context, the $5d$ electron systems are particularly well suited because the strength of the Coulomb interaction and the spin-orbit coupling are of similar orders of magnitude \cite{5dsystem}, possibly giving rise to novel electronic phases such as a magnetic Weyl semimetal \cite{WSM}, a topological Mott insulator \cite{TMI}, or a Weyl Mott insulator \cite{WMI} state. Moreover, non-collinear or non-coplanar spin textures in chiral magnets have been shown to induce a large Berry curvature, leading to the appearance of topological or unconventional anomalous Hall effect (AHE) \cite{PIO2, Nd2Mo2O7, MnSi, Mn3Sn}.

Pyrochlore iridates $R_2$Ir$_2$O$_7$ (where $R$ is a lanthanoid or yttrium) are interesting from both points of view, having an iridium $5d$ electron system and non-collinear or non-coplanar spin texture due to the geometrically frustrated magnetism in both $5d$ and $4f$ electron sectors \cite{PIO3, StrainedFilmTheory, 5dsystem, PressureEIO, Tian, Ma, PIO2, NewCriticality2, PIO1, PIO4, ARPES, Quadratic, MIT}. The electronic transport of pyrochlore iridates varies from metallic to nonmetallic as the $R^{3+}$ ionic radius decreases \cite{5dsystem, MIT}. A metal-insulator transition occurs in almost all compounds in this series, except for $R$ = Pr, which has the largest ionic radius: Pr$_2$Ir$_2$O$_7$ remains metallic down to the lowest experimental temperatures \cite{PIO1}. Moreover, unconventional magnetotransport is observed due to the coupling between the $5d$ conduction electrons and the frustrated $4f$ moments on the pyrochlore lattice \cite{PIO2, PIO1, PIO3, PIO4}. The Pr $4f$ moments are located at the vertices of the pyrochlore lattice with an Ising anisotropy along the $\langle111\rangle$ direction. No long range magnetic order is observed down to the lowest measurement temperature as a ferromagnetic coupling ($J \sim 0.7$ K) between the Ising $4f$ moments leads to a frustrated spin ice state with a local 2-in--2-out structure \cite{PIO3, PIO1}. Furthermore, the system shows a spontaneous Hall effect, i.e., AHE without spontaneous magnetization or external magnetic field between 0.3 K and 1.5 K \cite{PIO2, PIO3}. The AHE is most likely topological, originating in the spin chirality generated by a 2-in--2-out non-coplanar spin ice configuration below the spin ice correlation scale of $2J \sim 1.5$ K. This spin configuration is consistent with the observed anisotropic magnetotransport, i.e., the non-zero remnant Hall resistivity that reaches a maximum when a magnetic field is applied along the [111] direction \cite{PIO4}, indicating that the cubic symmetry should be broken likely by the scalar chirality order of $4f$ moments in the spin liquid state.

More recently, paramagnetic band calculations suggest that Pr$_2$Ir$_2$O$_7$ has a Fermi node, formed by quadratic band touching of the doubly degenerate valence and conduction bands at the $\Gamma$ point at the Fermi level ($E_{\rm F}$) \cite{Quadratic, ARPES}. In the stoichiometric Pr$_2$Ir$_2$O$_7$ system, the charge neutrality locates $E_{\rm F}$ exactly at the node but any off-stoichiometry may dope electrons (holes) in the conduction (valence) band. The presence of the Fermi node near the Fermi energy has been experimentally observed by angle-resolved photoelectron emission spectroscopy \cite{ARPES}. This nodal state indicates that Pr$_2$Ir$_2$O$_7$ is sensitive to perturbations and may host various topological phases \cite{Quadratic, ARPES}. In particular, it has been predicted that by breaking the cubic symmetry and/or TRS, the nodal state can be converted to a topological insulator or a Weyl metal state \cite{Quadratic, ARPES}. In contrast to vigorous theoretical studies, experimental observations of the topological phases developed from the nodal state remain elusive.

To date the experimental studies on Pr$_2$Ir$_2$O$_7$ have been limited to bulk samples. Thin films offer a suitable platform for additional control over the crystal growth orientation and lattice deformation by compressive or tensile strain imposed by epitaxial lattice mismatch with a substrate. In this work, (111)-oriented pyrochlore Pr$_2$Ir$_2$O$_7$ thin films were epitaxially grown on yttria-stabilized zirconia (YSZ) (111) substrates. As the lattice constant of bulk Pr$_2$Ir$_2$O$_7$ (10.394 \AA~\cite{PIO1}) is larger than that of YSZ ($2a=10.278$~\AA), an epitaxial thin film may be expected to be compressively strained biaxially in the in-plane direction, which leads to tensile strain along the surface normal [111] direction (Experimental).

\section{Experimental}
\subsection{Sample fabrication}
Pyrochlore Pr$_2$Ir$_2$O$_7$ thin films were fabricated on YSZ(111) single crystal substrates using pulsed laser deposition (PLD) at room temperature followed by solid phase epitaxy. Prior to deposition, the YSZ substrates were annealed in an electric furnace at 1250 $^\circ$C for 2 hours in air to obtain atomically flat step-and-terrace surface \cite{YSZ}. The PLD target was prepared by mixing polycrystalline Pr$_2$Ir$_2$O$_7$ and IrO$_2$ powders, giving a 1:2 Pr/Ir ratio to compensate for the loss of Ir in the PLD process due to the high volatility of Ir oxides \cite{IrVolatility}. The mixed powder was compacted into a pellet by spark plasma sintering (SPS) at 900 $^\circ$C for 10 minutes under an applied pressure of 100 MPa in an Ar atmosphere of 0.03 MPa. A KrF excimer laser (Tui Laser, $\lambda$ = 248 nm) was used to ablate the sintered Pr-Ir-O target. The laser fluence and repetition rate were set to 0.65 J/cm$^2$ and 5 Hz, respectively. The deposition was done at room temperature at an oxygen pressure of 10 mTorr. After deposition, the films were crystallized by post-annealing in an electric furnace at 1000 $^\circ$C for 1.5 hours in air. The film thickness was 100 nm as measured with a stylus profilometer (DEKTAK, Veeco).

\subsection{Characterization}
The crystal structures of the samples were analyzed by XRD (SmartLab, Rigaku). Hall bars for transport measurements were fabricated by mechanical diamond milling. Microwedge samples for high-angle annular dark field scanning transmission electron microscopy (HAADF-STEM) observation were cut from the central parts of the Hall bars by focused ion beam (FIB) milling after transport analysis. Resistivity and magnetotransport measurements were done in a physical property measurement system (PPMS, Quantum Design) and a top-loading dilution refrigerator (Kelvinox, Oxford) above and below 1 K, respectively. For measurements below 1 K, the sample was directly loaded into the cryostat mixing chamber to ensure a homogeneous temperature at the film surface. The measurements were performed by standard low-frequency (13 Hz) ac lock-in technique with an excitation current of 10 nA (Fig.~\ref{figure2}(b)) or 100 nA (Figs.~\ref{figure3}(a), \ref{figure3}(b), and \ref{figure3}(d)). Magnetization curves were obtained by using a superconducting quantum interference device magnetometer (MPMS, Quantum Design). The substrates did not affect the transport and magnetic measurements because YSZ is a good insulators and nonmagnetic.

\section{Results}
\subsection{Crystal structure analysis}

\begin{figure*}
\includegraphics[width=16cm, clip]{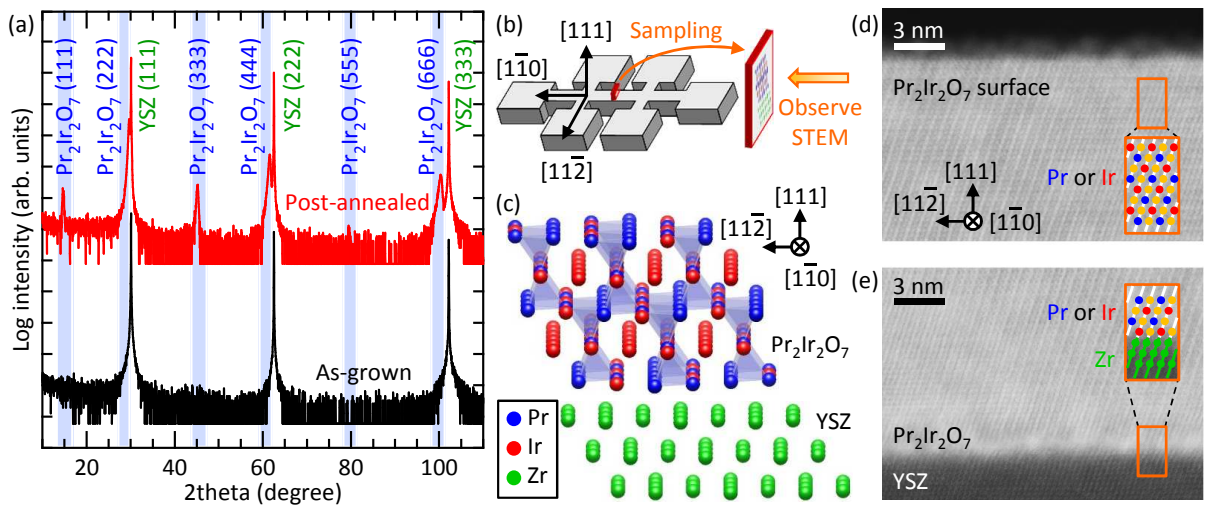}%
\caption{\label{figure1}Crystal structure analysis of Pr$_2$Ir$_2$O$_7$/YSZ(111). (a) $\theta$--$2\theta$ XRD patterns of as-grown (bottom) and post-annealed (top) Pr-Ir-O thin film on a YSZ(111) substrate. Thin film peaks are highlighted in blue. (b) Schematic illustration of STEM sampling and observation direction. A microwedge sample was cut from the central part of the Hall bar by focused ion beam milling. (c) Cross-sectional crystal structure of the Pr$_2$Ir$_2$O$_7$/YSZ(111) interface \cite{VESTA}. Pr, Ir, and Zr atoms are drawn in blue (\textcolor[rgb]{0,0,1}{\textbullet}), red (\textcolor[rgb]{1,0,0}{\textbullet}), and green (\textcolor[rgb]{0,1,0}{\textbullet}), respectively. Oxygen is omitted. The model is slightly tilted around the $[11\overline{2}]$ direction to show the atomic arrangement in the depth direction. (d) and (e) Cross-sectional HAADF-STEM images of a single grain in the film, taken near the Pr$_2$Ir$_2$O$_7$ surface and the Pr$_2$Ir$_2$O$_7$/YSZ interface, respectively. Pr, Ir, and Zr atom positions are shown in the insets for the regions marked with the orange outlines. The colors correspond to those used in (c). Yellow (\textcolor[rgb]{1,0.75,0}{\textbullet}) represents Pr and Ir atoms that are alternately arranged in the depth direction. Crystal axes are shown in (d). The scale bars correspond to 3 nm.}
\end{figure*}

Figure \ref{figure1}(a) shows x-ray diffraction (XRD) patterns of as-grown and post-annealed Pr-Ir-O/YSZ(111) films. The annealing procedure after deposition crystallizes the as-deposited amorphous film, leading to the formation of epitaxial Pr$_2$Ir$_2$O$_7$, which can be identified by the odd-numbered pyrochlore (111) peaks. Additional characterization of the epitaxial growth and lattice relaxation are described in Supplemental Fig.~S1 \cite{SupplementalMaterial}. Although the XRD analysis showed that the film is epitaxial, reciprocal space analysis indicated that the film is mostly relaxed. On the other hand, microscopic cross-sectional imaging of individual grains in the Pr$_2$Ir$_2$O$_7$/YSZ(111) film by HAADF-STEM reveals that strained grains exist in the film. Figures \ref{figure1}(b) and \ref{figure1}(c) show the STEM observation direction and a lattice model viewed along the STEM electron beam direction. The Pr and Ir atoms in the film and the Zr atoms in the substrate are responsible for the periodic contrast in Figs.~\ref{figure1}(d) and \ref{figure1}(e) since the intensity of a column of atoms in a HAADF-STEM image is proportional to the average atomic number in that column. This effect can be seen in the film part of the STEM images, where $^{77}$Ir (red) columns are brighter than the $^{59}$Pr (blue) columns. The contrast between the two columns in the STEM image shows that the atomic arrangement in the film matches the expected atomic arrangement of an ordered pyrochlore lattice illustrated in Fig.~\ref{figure1}(c). The insets of Figs.~\ref{figure1}(d) and \ref{figure1}(e) show diagonal parallel lines connecting neighboring atom positions. At the interface, each line in the film continuously extends into the substrate, showing that the atom row spacings at the interface and the grain surface are identical and there are no misfit dislocations at the interface in the STEM imaging area. The STEM image thus proves that at least some grains in the Pr$_2$Ir$_2$O$_7$ film are locked to the YSZ substrate and not fully relaxed, as is also suggested by the peak shoulder in the XRD reciprocal map (Supplemental Fig.~S1 \cite{SupplementalMaterial}). Such locked grains are coherently grown from the interface to the grain surface. The lattice mismatch of the film relative to YSZ is +1.15\%, which corresponds to a tensile strain of 2.31\% along the [111] direction when the film lattice is in-plane locked to YSZ.

\subsection{Temperature dependence of the longitudinal resistivity}
\begin{figure}
\includegraphics[width=8.6cm, clip]{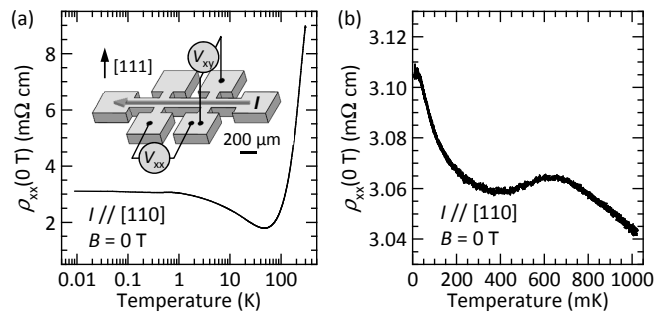}%
\caption{\label{figure2}Temperature dependence of the longitudinal resistivity of a Pr$_2$Ir$_2$O$_7$ thin film. (a) Plot over a wide temperature range, measured at zero magnetic field. Inset shows a schematic illustration of the Hall bar geometry. The current flows along the [110] direction. (b) Enlargement of (a) below 1 K.}
\end{figure}

Figure \ref{figure2}(a) displays the temperature dependence of the longitudinal resistivity of a Pr$_2$Ir$_2$O$_7$ thin film measured in zero magnetic field ($\rho_{{\rm xx}}(0~{\rm T})$). The resistivity decreases monotonically as the film is cooled down from room temperature, indicating metallic conductivity. The carrier density at 2 K is estimated to be 1.75$\times10^{20}$ cm$^{-3}$, suggesting that a slight off-stoichiometry causes hole doping in the valence band and shifts $E_{\rm F}$ by 18 meV (see Supplemental Fig.~S2 \cite{SupplementalMaterial}). A resistivity minimum appears at 47 K, followed by a nondivergent upturn when approaching 0 K. The behavior is similar to the one reported for bulk Pr$_2$Ir$_2$O$_7$ \cite{PIO1}. However, a closer look at the lowest temperature region in Fig.~\ref{figure2}(b) shows that there is a small suppression below about 700 mK. This anomaly can be attributed to Q = (001) antiferromagnetic order of the Pr $4f$ moments \cite{AntiOrder} as discussed later.

Compared to the single crystalline bulk case \cite{PIO1}, the absolute value of $\rho_{{\rm xx}}(0~{\rm T})$ is about one order of magnitude higher in the thin film, mostly due to grain boundary scattering. Although off-stoichiometry due to Ir volatilization is conceivable, it has been reported that an Ir loss of more than a few percent drastically changes the temperature dependence of $\rho_{{\rm xx}}(0~{\rm T})$ \cite{Prstuffing}. Specifically, the temperature of the resistivity minimum shifts to 20 K in the Pr-rich phase and the resistivity behavior below 4 K is quite different from the stoichiometric case, showing a sharp metallic decrease due to a large number of doped carriers, rather than the nearly temperature-independent behavior seen here. It is therefore most likely that the uniform increase of $\rho_{{\rm xx}}(0~{\rm T})$ is an intrinsic property of the nearly stoichiometric Pr$_2$Ir$_2$O$_7$ film.

\begin{figure}
\includegraphics[width=8.6cm, clip]{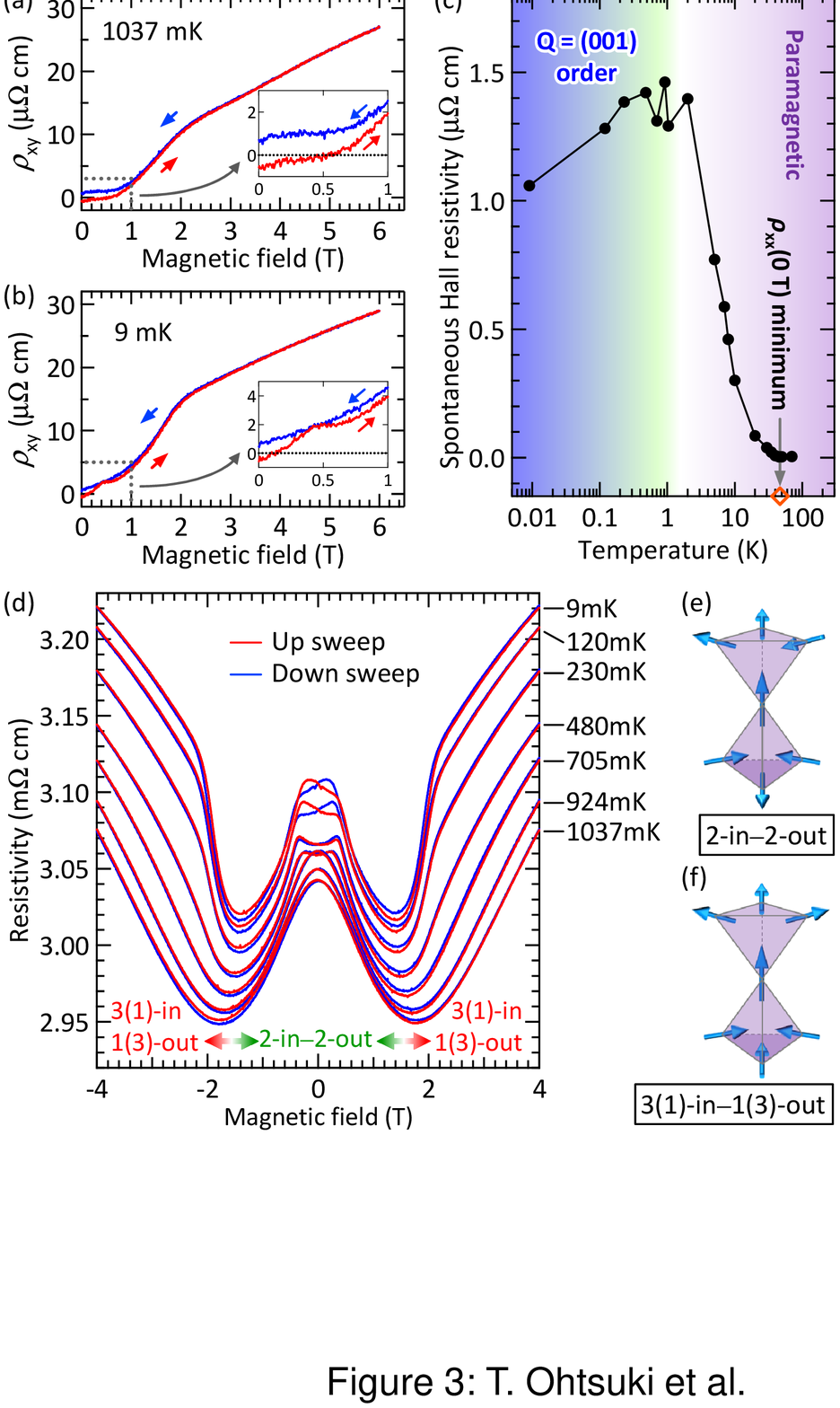}%
\caption{\label{figure3}Magnetotransport properties of a Pr$_2$Ir$_2$O$_7$ thin film. (a) and (b) Magnetic field ($B$) dependence of the Hall resistivity ($\rho_{{\rm xy}}(B)$), measured at 1037 mK and 9 mK, respectively. $\rho_{{\rm xy}}(B)$ is defined as $[\rho_{{\rm xy}}(B)-\rho_{{\rm xy}}(-B)]/2$ to eliminate the $\rho_{{\rm xx}}$ component. The insets show the magnified plot for the low field hysteresis part of $\rho_{{\rm xy}}(B)$. (c) Temperature dependence of the spontaneous Hall resistivity, i.e., the absolute value of $\rho_{{\rm xy}}(B)$ at $B=0$ obtained after a field cycle. Orange open diamond (\textcolor[rgb]{1,0.33,0}{$\diamond$}) indicates the temperature at which $\rho_{{\rm xx}}(0~{\rm T})$ reaches a minimum. (d) Transverse MR curves as a function of magnetic field. Measurement temperature are denoted on the right. Spin configurations labeled in (c) and (d) are indicated for Pr $4f$ moments. The magnetic field was applied along the [111] direction in all measurements. Red and blue lines in (a), (b), and (d) corresponds to up and down sweeps of the magnetic field, respectively. Illustrations of the (e) 2-in--2-out and (f) 3(1)-in--1(3)-out spin configurations for Pr $4f$ moments.}
\end{figure}

Similar features in the temperature dependence of $\rho_{{\rm xx}}(0~{\rm T})$, i.e., a minimum and nondivergent upturn towards 0 K have been observed in other metallic pyrochlore iridates \cite{PressureEIO, Tian, PIO1}. However, the origin is still controversial. One possibility is the Kondo effect, as discussed in the context of bulk Pr$_2$Ir$_2$O$_7$ \cite{PIO1} for the interaction of Ir $5d$ conduction electrons with the localized Pr $4f$ magnetic moments. In contrast, theoretical calculations have also suggested the possibility of a non-Kondo mechanism \cite{Rminimum}. In this scenario, conduction electrons are elastically scattered by a geometrically frustrated spin-ice-like correlation that develops as the temperature is lowered. The computed results have been compared with experimental data on Pr$_2$Ir$_2$O$_7$ \cite{PIO1}, reproducing the main features of the resistivity, specific heat, and magnetic susceptibility. Finally, the resistivity upturn may be related to the quadratic band touching that is a unique characteristic of the Pr$_2$Ir$_2$O$_7$ band structure \cite{Quadratic, ARPES}.

\subsection{Hall effect measurements}
The most striking feature was found in the magnetic field ($B$) dependence of the Hall resistivity ($\rho_{{\rm xy}}(B)$) shown in Figs.~\ref{figure3}(a) and \ref{figure3}(b) (Supplemental Fig.~S3 for the data above 2 K \cite{SupplementalMaterial}). Similarly to the bulk case, $\rho_{{\rm xy}}(B)$ exhibits a hysteresis loop near zero field. However, a non-zero spontaneous Hall resistivity at zero field, $\rho_{{\rm xy}}(0)$, persists up to $T_{\rm H} \sim 50$ K, much higher than the onset temperature ($T_{\rm H} \sim 1.5$ K) for the bulk. As shown by the temperature dependence plot in Fig.~\ref{figure3}(c), $T_{\rm H} \sim 50$ K coincides with the $\rho_{{\rm xx}}(0~{\rm T})$ minimum temperature, below which the spontaneous Hall resistivity increases down to 700 mK. A slight reduction is seen below 700 mK most likely due to the Q = (001) order that sets in at around 700 mK, as will be discussed below.

In contrast, no hysteresis was found in the magnetization curve at an experimental resolution of $\sim$0.001 $\mu_{\rm B}$. A typical result at 2 K is shown in Supplemental Fig.~S4 \cite{SupplementalMaterial}. Thus, our Pr$_2$Ir$_2$O$_7$ thin film exhibits a spontaneous Hall effect without magnetic field and at zero magnetization within our experimental accuracy. Thus, the spontaneous Hall effect should not be caused by the conventional AHE related to ferromagnetism, but by the unconventional one due to a non-coplanar or non-collinear spin texture, namely, a spin liquid state or antiferromagnetic order, similarly to bulk Pr$_2$Ir$_2$O$_7$ \cite{PIO3}. Interestingly, the onset temperature scale, $T_{\rm H} \sim 50$ K, is too high for the exchange coupling ($J \sim 0.7$ K) among Pr $4f$ moments. Thus, the spontaneous AHE should most likely come from the magnetic order formed by Ir $5d$ electrons. Possible appearance of a magnetic order has been theoretically predicted for pyrochlore iridate thin films \cite{OrderThinFilm2}.

\subsection{Transverse magnetoresistance} We now turn to the low temperature magnetism due to Pr $4f$ moments. The magnetotransport characteristics of the Pr$_2$Ir$_2$O$_7$ thin film are shown in Fig.~\ref{figure3}(d). The transverse magnetoresistivity (MR) decreases initially as the magnetic field increases from 0 T, indicating that magnetic moments become partially aligned along the magnetic field, [111] direction (Fig.~\ref{figure3}(d)). The negative MR only appears below 50 K, where $\rho_{{\rm xx}}(0~{\rm T})$ passes a minimum (Supplemental Fig.~S5 \cite{SupplementalMaterial}). With further increasing field, the MR becomes positive across a field $\sim$1--2 T. Based on bulk Pr$_2$Ir$_2$O$_7$ magnetization measurements \cite{PIO3}, this crossover field can be associated with a metamagnetic transition from a 2-in--2-out (Fig.~\ref{figure3}(e)) to a 3(1)-in--1(3)-out (Fig.~\ref{figure3}(f)) configuration of Pr $4f$ moments below the spin ice correlation scale of $2J \sim 1.5$ K, and with the field polarization of the spins at high temperatures. These are consistent with the linear temperature dependence of the crossover field (Supplemental Fig.~S6 \cite{SupplementalMaterial}). Close to the crossover field, the MR passes through a minimum and shows a small hysteresis between the up and down sweeps below 2 K.

Below 700 mK where an anomaly was observed in the temperature dependence of $\rho_{{\rm xx}}(0~{\rm T})$ (Fig.~\ref{figure2}(b)), the MR further exhibits an additional kink accompanied by a hysteresis loop around 0 T. This zero field anomaly is very close to the kink found in the bulk caused by the Q = (001) antiferromagnetic order \cite{AntiOrder}. The fact that this phase can be easily suppressed by a weak field is consistent with the antiferromagnetic ordering.

\subsection{Magnetic phase diagram due to Pr 4$f$ moments} 
\begin{figure}
\includegraphics[width=8.6cm, clip]{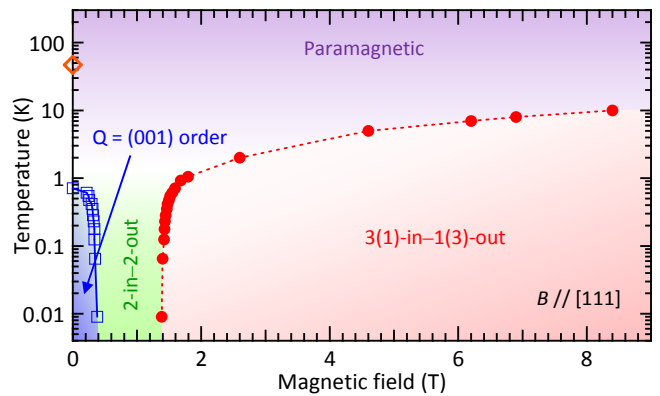}%
\caption{\label{figure4}Magnetic phase diagram for a Pr$_2$Ir$_2$O$_7$ thin film as a function of temperature and magnetic field, mainly governed by Pr 4$f$ moments. The Q = (001) order, 2-in--2-out and 3(1)-in--1(3)-out states appear in areas colored in blue, green, and red, respectively. Blue open squares (\textcolor[rgb]{0,0,1}{$\square$}) denote the magnetic field values below which a hysteresis loop is observed in the vicinity of zero magnetic field in the MR curves in Fig.~\ref{figure3}(d). Red solid circles (\textcolor[rgb]{1,0,0}{\textbullet}) represent points where the MR curves in Fig.~\ref{figure3}(d) and Supplemental Fig.~S5 pass through a minimum. Orange open diamond (\textcolor[rgb]{1,0.33,0}{$\diamond$}) indicates the temperature at which the spontaneous Hall effect starts to develop. The magnetic field was applied along the [111] direction.}
\end{figure}

Summarizing the transport measurements, we construct a magnetic phase diagram mainly governed by Pr $4f$ moments (Fig.~\ref{figure4}). The Q = (001) order most likely appears only in the low-field and low-temperature regions and can be suppressed by either magnetic field or temperature. The low-field anomaly in MR is due to the suppression of the Q = (001) order in a magnetic field. At low temperatures, the crossover from the 2-in--2-out to the 3(1)-in--1(3)-out configuration is visible as the MR minimum. As expected, the data points on this crossover curve have a linear relationship between temperature and magnetic field (Supplemental Fig.~S6 \cite{SupplementalMaterial}).

\section{Discussion}
Finally, we discuss the origin of the spontaneous Hall effect and its possible relation to a Weyl semimetal phase. Theoretically, in a pyrochlore iridate with a quadratic band touching such as in the present Pr$_2$Ir$_2$O$_7$ system, both the cubic symmetry and the macroscopic TRS must be broken for a Weyl semimetal phase to appear \cite{HallinWeyl, Quadratic}. The observation of the spontaneous Hall effect provides strong evidence for the macroscopic TRS breaking. As we discussed, the temperature scale ($T_{\rm H} = 50$ K) where the spontaneous non-zero Hall resistance appears is too high for Pr moments to form a long range magnetic order or a chiral spin liquid phase. Therefore, it is most likely that the Ir $5d$ electrons are responsible for the TRS breaking. Recent calculations suggest that thin films may show a magnetic order near the surface or at an interface as the nearest neighbors are lost and the band becomes narrower \cite{OrderThinFilm2}. The most likely spin configuration is all-in--all-out, as this is the only spin order reported to date among pyrochlore iridates \cite{Neutron, IrOrdering1, IrOrdering2} and Nd$_2$Ir$_2$O$_7$, which has a slightly smaller band width than Pr$_2$Ir$_2$O$_7$, shows this type of spin order \cite{Neutron}. However, for the system to break the macroscopic TRS, the all-in--all-out state by itself is not enough, and further breaking of the cubic symmetry is required. We thus argue that both preconditions for the spontaneous Hall effect to appear, i.e., the all-in--all-out spin configuration and strain along the [111] direction are fulfilled in the strained parts of the film, as has been pointed out in Ref.~\cite{HallinWeyl, Quadratic}. We therefore conclude that these experimental results are consistent with the appearance of the Weyl semimetal phase in Pr$_2$Ir$_2$O$_7$ films.

As was discussed earlier, the onset of another magnetically ordered state is observed below 700 mK. This most likely comes from the Q = (001) order of Pr $4f$ moments, as seen in bulk samples \cite{AntiOrder}. Since this state does not carry a net magnetization and does not macroscopically break TRS, an increase in the spontaneous Hall effect would not be expected. Moreover, this ordered state of Pr $4f$ moments is stabilized by ferromagnetic (RKKY type) coupling mediated through Ir $5d$ bands and thus the phenomenon should originate in the relaxed bulk part of the film and not the strained grains where antiferromagnetic correlation stabilizes the all-in--all-out order as discussed above. However, as the bulk part and the strained part are connected physically, the two interactions may compete to suppress the spontaneous Hall effect. This would be the reason why a slight suppression of the spontaneous Hall resistivity is found below 700 mK. Our study demostrates that the thin film fabrication provides a significant route to explore the novel topological phases in correlated oxides.

\section{Acknowledgments}
We are grateful to Prof.~K.~Kimura (Univ.~of Tokyo) and Dr.~Y.~Takagiwa (National Institute for Material Science) for help with the SPS processing. This work was supported by CREST (Grant No. JPMJCR15Q5), Japan Science and Technology Agency, Grants-in-Aid for Scientific Research (Grant Nos. 16H02209, 25707030 and 26105002), by Grants-in-Aid for Scientific Research on Innovative Areas ``J-Physics" (Grant Nos. 15H05882 and 15H05883), Program for Advancing Strategic International Networks to Accelerate the Circulation of Talented Researchers (Grant No. R2604) from the Japanese Society for the Promotion of Science, and the support of ICAM Branches Cost Sharing Fund.


\end{document}